\documentclass[prl,multicol,epsf,epsfig,amsmath]{revtex4}
\usepackage[toc,page]{appendix}
\usepackage{epsfig}
\newcommand{\beq}{\begin{equation}}

\newcommand{\va}{\varepsilon}
\newcommand{\eeq}{\end{equation}}
\newcommand{\barray}{\begin{eqnarray}}
\newcommand{\earray}{\end{eqnarray}}
\newcommand{\nn}{\nonumber}

\newcommand{\eq}[1]{Eq.~(\ref{#1})}

\newcommand{\am}{{\bf a}}
\newcommand{\A}{{\bf A}}
\newcommand{\bm}{{\bf b}}
\newcommand{\B}{{\bf B}}
\newcommand{\Sm}{{\bf S}}
\newcommand{\num}{{\bf { \cal V}}}

\begin{document}
\title{Parameter Dependent Commuting Matrices, Pl\"ucker relations  and Related  Quantum Glass Models }
\author{B Sriram Shastry }
\address{Department of Physics, University of California, Santa Cruz, CA 95064}
\date{\today}
\begin{abstract}
Type-I matrices were introduced recently as finite dimensional prototypes of quantum integrable systems. These matrices are linearly dependent on an ``interaction''  type  parameter, and possess interesting properties such as commuting partner matrices and generically violate the von Neumann Wigner non crossing rule.
The important role of Pl\"ucker relations in this construction is noted. 
 Type-I matrices are given a transparent formulation in terms of Fermi or Bose type particle operators- they represent  a Quantum glass model with either Fermi or Bose statistics, with several free parameters that may be chosen at will.
\end{abstract}
\maketitle
\section{Introduction }
In Ref \cite{shastry_05} (referred to as I below),  the author initiated a study of quantum integrable systems  in finite dimensions, within the context of parameter dependent commuting matrices. This in turn was motivated by several examples of specific integrable models, such as the Hubbard model\cite{heilmann-lieb,shastryrrr,grabowski} and the Heisenberg model \cite{heisenberg}. In these examples,  one studies the  realizations of the general model in Fock space for particular sectors of quantum numbers, such as momentum, parity, total spin  and number. This lead one to  real symmetric matrices in various dimensions $N \geq 2$. These   have the remarkable feature that   the  von Neumann Wigner non crossing rule\cite{vnw} is  violated.  One ends up with several  level crossings that are conventionally termed ``accidental''.  This terminology is rather avoidable,  since there is a also a belief that there is nothing accidental in having such level crossings;  the existence of several dynamical conservation laws (dependent on coupling constants)  are believed to be causally implicated. 
Further, the statistics of energy levels of these integrable models are also known to be close to Poisson statistics, and hence consistent with the absence of level repulsion that generic systems are known to possess\cite{mehta,poilblanc}. 

While the current  general programme for the study of quantum integrable systems focus on properties such as factorizable S matrices, or the Yang Baxter relation, the approach of (I)  Ref \cite{shastry_05}  gets to the core of the issue of the matrix realizations of these models.  As such, it  is ``blind'' to the specific physical details of the models. The main results of (I)  \cite{shastry_05} is the identification of a class of matrices, termed Type-I matrices, discussed in detail below. Here the core property of multiple parameter dependent conservation laws is made explicit, and one has an algorithm for generating such matrices as well as a count of the number of such matrices.

In an impressive work,   Ref.\cite{Owusu} Owusu, Wagh and Yuzbashyan  (OWY) have built on this initial advance, and produced  several further  results. A fundamental advance  is the introduction of a basis of matrices, in terms of which the matrices of Ref. \cite{shastry_05} can be expanded. OWY   further show  a link with an integrable model due to Michel Gaudin\cite{gaudin}, that is currently enjoying popularity in the context of superconductivity of finite systems\cite{gaudin}.  OWY also throw light on  the ``mechanism''  of the the level crossing, and give explicit formulas for the number of level crossings one finds in Type-I matrices.

The objective of this paper is multifold. Firstly,  a fundamental constraint equation in  Ref.\cite{shastry_05} for constructing Type-I matrices  is shown to be  related to the so called Pl\"ucker relations of Pfaffians. Since Pfaffians are basic  to anticommuting objects such as Majorana Fermions,  one sees that Fermi statistics enters this program of describing integrable systems in a fundamental and  unexpected fashion. 
 From this analysis, the   parameterization of the solutions of  Type-I matrices   by OWY in  Ref.\cite{Owusu},  arises as an elegant  consequence,  and the entire construction becomes more transparent.

 Secondly, I show that the link with the Gaudin  type model\cite{gaudin} is made more naturally with Fermi (or canonical Bose) statistics.  The connection made by OWY with the Gaudin model assumed hard core Bose statistics for the particles, and is  confined to   the sector of one spin wave, i.e. is confined to a specific sub manifold. The basic quantum operator underlying this class of problems is the permutation operator that has several possible realizations, leading to distinct models. The permutation operator  has a Fermi representation: this  is shown to be more natural than the (hard core) Bosonic one used in Ref.\cite{Owusu}.  Once the commutation relations  of a set of  matrices is established, we can elevate these to operator relations with either Fermi or Bose statistics (see Eq.~(\ref{elementary}) below) and thus also have a Bosonic representation of these. 
 
 With either Fermi, or with soft core (i.e. canonical) Bosons, we construct a  Quantum glass model below, i.e. a particular type of Anderson model for disordered carriers. This model is akin to Gaudin's model with hard core spins, and depends on several parameters that may be chosen as one wishes, and  has commuting partners in {\em all particle sectors}. These commuting partners may be thought of as  local charges  that are broadened out in a specific way.

Interestingly  the Pl\"ucker relations arise  in other aspects of integrable systems as well. 
these are  central objects in Sato's work on classical solitonic theories  (i.e. classically integrable systems)\cite{sato_1,sato_2}, where the so called $\tau$ functions satisfy these relations. 
For  quantum integrable models,  a connection has been shown to exist between the transfer matrices of and the bilinear identities of the $\tau$ functions\cite{weigmann} satisfying Pl\"ucker relations.

 A few remarks  are useful to put  the current work and the related Refs.\cite{shastry_05,Owusu},   within the  context  of  matrix theory as used in Quantum theory.  In order to keep things simple, let us specialize to finite dimensions \footnote{ By sticking to lattice models in finite dimensions, we are following the    Marc Kac  dictum: {\em \\ Be Wise, \\ Discretize.}}.
Quantum observables lead to Hermitean, or real symmetric matrices, and simultaneous measurability of  observable   pairs translates to the theorem that commuting matrices of the above type are simultaneously diagonalizable. One simple result about two such  commuting matrices $a$ and $b$, is that one of them is expressible as a power series in the other \footnote{If one of them $a$  has  distinct eigenvalues, say $a_{\lambda}$, then it is possible  to express  $b= \sum_{j=1}^{N}c_{j}a^{j}$ with suitable constants $c_{j}$, and $N$ is the dimension of the matrices. This expansion is most easily seen in the basis where both matrices are diagonal. For the linear equations to be consistent,  the Vandermonde determinant   $\prod_{\lambda < \mu}(a_{\lambda}-a_{\mu}) \neq 0$  is required to be non vanishing,  leading to the requirement of  non degenerate eigenvalues. The case of degeneracy is obtained by taking suitable limits within this framework.}. The  current series of works differ from these in that the focus is on matrices that depend in a simple way (linearly, or possibly a polynomial of low degree) on a parameter, and one insists upon the commutation property for all values of the parameter. This problem is natural in the context of examples in Quantum theory, such as the hydrogen atom, where the Laplace Runge Lenz vector   depends linearly on the squared electric charge. It is  also true in the structure of the {\em higher conservation laws} in models such as the Heisenberg \cite{heisenberg} and Hubbard models\cite{shastryrrr,grabowski}. 
The notable results in the works  Refs.\cite{shastry_05,Owusu}
follow from  the detailed 
study of the simple parameter  dependence  of the commuting pair.

\section{Summary of (I) and the introduction of a basis of commuting operators }
In Ref \cite{shastry_05}, 
we introduced a family of real symmetric matrices  in $N$ dimensions depending linearly on a parameter $x$. These were introduced as purely algebraic prototypes of integrable systems in finite dimensions and termed as Type-I matrices. They are efficiently represented as:
\beq
{\bf \alpha} = \am + x \A, \;\; \; \mbox{with}\;\;\; \A= {\bf A}_{d}+  [\am, \Sm], \label{eq1}
\eeq
with  two  generic diagonal matrices  $\am$ and ${\bf A}_{d}$ having unequal entries, i.e.   $\am= DiagonalMatrix\{u_{1},\ldots,u_{N}\} $  with $u_{i}\neq u_{j}$\footnote{We have changed the notation here from \eq{eq1} and \eq{eq2} with  $a_{r}\to u_{r}$ and $b_{r}\to v_{r}$ in order to avoid a conflict with the notation of the Fermionic operators $a_{r},a^{\dagger}_{r}$.},  ${\bf A}_{d}= DiagonalMatrix\{A_{1},\ldots,A_{N}\} $  with $A_{i}\neq A_{j}$,
and  a real antisymmetric matrix $S_{i,j}=-S_{j,i}$.  One may think of $\am$ and $\A$ as the kinetic and potential energy matrices, 
and the parameter $x$  as a perturbation parameter
 in typical quantum systems. In this notation, the role of the antisymmetric matrix $S_{ij}$ is made explicit.  There is no loss of generality since if we are given the matrix $\alpha(x)$ in an arbitrary basis as the sum of two non commuting matrices, we can convert it to this form by performing an orthogonal transformation that diagonalizes the matrix $\alpha(x= 0)$.
 
 In this way, we  model integrable systems, without reference to their explicit origin in the physical world, as parameter dependent matrices. This construction is inspired by the  standard examples of the Hubbard and Heisenberg models. In these models,  finite dimensional matrices of the above type emerge on restricting the state space to various sectors of usual  (parameter independent)  conservation laws such as  particle number, parity, spin and total momentum.
 
Since integrable systems are known to  possess several parameter dependent (i.e. dynamical) conservation laws, one wants to know if other matrices depending on $x$, possibly linearly, can be found. It was indeed shown that  under certain conditions on S,  summarized below,  such commuting partners $\beta(x)$ can be found, i.e. $ [{\bf \alpha,\beta }]=0$.
The form of the dynamical conservation laws $\beta(x)$ was shown to be very similar as that of  $\alpha(x)$:
\beq
{\bf \beta} = \bm + x \B, \;\;\; \mbox{with}\;\;\; \B= {\bf B}_{d}+ [\bm,\Sm], \label{eq2}
\eeq
where $\bm$ and ${\bf B}_{d}$  are   diagonal matrices.

 In (I), it was shown that the number of 
 matrices of type-I  in $N$ dimensions is $\num_{a}= (3 N-1)$, and 
 for a given matrix $\alpha(x)$ out of this set, there are a further 
  $\num_{b}=N+1 $ matrices of the type $\beta(x)$. 
  
  The crucial condition on the antisymmetric matrix $\Sm$ was written in (I),  in terms of its inverse elements $R_{i,j}\equiv \frac{1}{S_{ij}}$ 
\barray
\phi_{(i,j,k,l)} &\equiv & R_{i,j} R_{k,l} - R_{i,k} R_{j,l} + R_{i,l}R_{j,k} = 0.  \label{rrrr}
\earray
 These equations are extensively discussed  in mathematics literature as the Pl\"ucker relations \cite{plucker}, and their analysis is presented later. In our original work (I), we noted that 
these are greatly overdetermined equations, since there 
are $^{N}C_4$ quartets of indices and equations, but only $^{N}C_2$ matrix elements $R_{i,j}$ to be determined. In (I) it was shown,  by using a consistency condition involving 5 indices(Eq(I-15)), that this set has $\num_{R}= 2 N-3$ free parameters and hence independent solutions. For example one may choose at will the parameters $R_{1,j}; \; 2\leq j \leq N$ and $R_{2,k};\;3\leq k \leq N$, and the remaining $R_{lm}$ are determined in terms of these with no conflicts.

In  Ref.\cite{Owusu} Owusu, Wagh and Yuzbashyan (OWY) have shown that it is more  efficient to introduce a  basis of commuting operators in terms of which both the matrices $\alpha$ \eq{eq1} and $\beta$ \eq{eq2} can be expanded linearly. Here  a commuting basis guarantees the commutation of the matrices $\alpha$ and $\beta$. The basis of    commuting operators $\{Z({r}) \}$, with $1 \leq r \leq N$  may be written in terms of the Dirac projection operators 
$
| i \rangle \langle j |,
$
as
\beq
Z({r})= | r \rangle \langle r | + x \sum_{s}\left[ \;  \rho_{s}(r) | s \rangle \langle s | + S_{r,s} \left( | r \rangle \langle s |+ | s \rangle \langle r | \right) \; \right]. \label{z-ops}
\eeq
Here $S_{r,s}$ will be seen below to be  the same elements as in \eq{eq1}.  The following  commutator  vanishes:
\beq
[Z({r}),Z({s})]=0, \label{z-commute}
\eeq
provided the matrix elements $\rho_{i}(j)$ satisfy the conditions
\barray
\sum_{r} \rho_{s}(r)&=& 0, \label{OWY-conditions-1} \\
\sum_{s} \rho_{s}(r)&=& 0, \label{OWY-conditions-3} \\
\frac{S_{i,l}S_{l,j}}{S_{i,j}}  &=&\Delta(i,j;l) =  \rho_{i}(l)-\rho_{j}(l). \label{OWY-conditions-2}
\earray
The constraint  \eq{OWY-conditions-3}  guarantees that the trace of the operators vanishes, it  is not necessary for the commutation of the $Z's$ but is a convenient condition.

 Now \eq{OWY-conditions-2} can be rearranged in a way that eliminates the $\rho_{i}(j)$ variables as a four index identity (where ``l'' is a spectator index)
\beq
\Delta(i,j;l)+\Delta(j,k;l)+\Delta(k,i;l)=0, 
\eeq
and in this form it is identical to Eq.~(I-10), and by using the inverse matrix elements  $R_{i,j}\equiv \frac{1}{S_{ij}}$, it becomes precisely \eq{rrrr} above.  The  point of this construction is that we can now take sums of the basis operators in Eq.~(\ref{eq1})
$$\alpha = \sum u_{j} Z({j}), \;\;\; \beta = \sum v_{j} Z({j})$$
and in this way recover the matrices found in (I).  By allowing some of the $a_{j}$ to be pairwise equal, OWY  obtain a somewhat greater freedom than that in (I), where all the $u_{j}$ were chosen to be  distinct in order to obtain  generic matrices. 

The number of  independent  parameters in the $S's$ or equivalently in $R's$ is $N (N-1)/2$,  and as mentioned above, we showed in (I) that the constraints in \eq{rrrr} are mutually consistent, giving  $2 N-3$ free parameters  in $S$.  OWY parameterize  the solutions  of \eq{rrrr} by  a neat  {\em ansatz}, namely 
\beq R_{i,j}= \frac{(\va_i-\va_j)}{(\gamma_i \gamma_j) }. \label{emil} \eeq
These are obtained in turn by  performing a local gauge transformation \eq{localgauge}, on a particular solution 
$ R_{i,j}= {(\va_i-\va_j)} $ noted in (I). This transformation consists of multiplying each matrix element by an arbitrary  $j$ dependent factor,
\beq
R_{i,j} \rightarrow 1/(\gamma_i \gamma_j) R_{i,j},  \;\; \phi_{ijkl}\to \phi_{ijkl} \ /(  \gamma_{i}\gamma_{j}\gamma_{k}\gamma_{l}) , \label{localgauge}
\eeq
and is clearly a way of generating further solutions from a given one.

  Indeed \eq{emil} has the correct number of parameters ($2 N-3$). To see this, we  start with 
the $N$ $\va's$, and  the $N$ $\gamma's$ giving us $2N$ parameters.  As discussed more fully below in \eq{scaling2},   we subtract 3 parameters from $2N$, since we can shift all $\va's$ by a single constant and further  scale all  the $\va's$ and all the $\gamma's$ by   two $j$ independent constants.
Using \eq{OWY-conditions-2},  we may then infer the $\rho_{j}(i)$ from this parametrization of $R_{ij}=1/S_{ij}$, and find 
\barray
\rho_{i}(j)&=& \frac{\gamma^{2}_{j}}{\va_{i}-\va_{j}}, \;\;\; i \neq j, \;\;\;\mbox{and} \nn \\
\rho_{i}(i)& = & \gamma^{2}_{i} \sum_{j\neq i}\frac{1}{\va_{i}-\va_{j}}. \label{rho-par}
\earray
We will use this convenient parameterization in the rest of this work. For completeness, we note that the parameterization of Eq.~(\ref{eq1}) i.e. $\alpha = \sum u_{j} Z({j})$ in (I) translates to the new variables as follows:
\barray
S_{ij}&=& \frac{\gamma_{i}\gamma_{j}}{\va_{i}-\va_{j}} \nn \\
A_{i}&=& \mbox{const}- \sum_{j \neq i} \gamma_{j}^{2} \ \frac{u_{i}-u_{j}}{\va_{i}-\va_{j}} \nn \\
Y_{ij}&=& \frac{\gamma^{2}_{i}}{\va_{i}-\va_{j}} - \sum_{k \neq i} \frac{\gamma^{2}_{k}}{\va_{i}-\va_{k}}, 
\earray 
with $Y_{ij}$ from Eq.~(I-5), so that $\mu(i;j k)$ in Eq.~(I-19)  vanishes identically.

\section{Pl\"ucker relations and the parametrization of the antisymmetric S matrix}
 \eq{rrrr} were  recognized belatedly by the author, as  Pl\"ucker relations of mathematical literature.    Since these  are central to the construction of this class of matrices we take a closer look at the $2N-3$ solutions that were found in (I).
We explore  the structure of the relations by using a more rigorous technique next, and see that the {\em ansatz} of OWY follows from the analysis as the unique solution.  We show  that these relations involve the so called 
Pl\"ucker relations for Grassman variables, and hence presage the final form our presentation that involves Fermions in a fundamental way.

We begin by noting that Eq(\ref{rrrr}) involves $\phi_{(ijkl)}$,
which is a  Pfaffian of a  real $4\times 4$ skew symmetric  matrix  $R_{i,j}$. 
The vanishing of $\phi_{(ijkl)}$ is a standard example of a Pl\"ucker relation\cite{plucker}.
The totality of these equations is expressed elegantly using
exterior forms. Let us define a $N$ dimensional real vector space spanned by unit vectors $e_j$ and define an antisymmetric  wedge products $e_i \wedge e_j$.  These provide a basis for the linear vector space $W(2)$\cite{plucker}. In this  space, 
 we define for a skew symmetric $R_{i,j}$ a ``two form'':
$$ {\cal R}= \sum_{i < j} R_{i,j} \ e_i \wedge e_j.$$ 
It is now easy  to see  that
$${\cal R }\wedge {\cal R}= 2 \sum_{i < j < k < l} \phi_{i,j,k,l} \; e_i \wedge e_j \wedge e_k \wedge e_l,$$
and hence we recognize that the totality  of  relations in \eq{rrrr}  are precisely  equivalent to finding solutions of   
\beq
{\cal R}\wedge {\cal R}=0. \label{wedge}
\eeq
This condition defines\cite{plucker} the ``decomposability'' of the two-form ${\cal R}$.
This problem can be resolved by noting that every skew symmetric matrix 
 can be expressed in  its real  normal form
 involving  $2 n$ orthonormal vectors $a^\alpha$ and $b^\alpha$ with $1\leq \alpha \leq n$,  satisfying the conditions
$\sum_j R_{i,j}a^\alpha_j= \lambda^\alpha b^\alpha_i$ and  $\sum_j R_{i,j}b^\alpha_j= - \lambda^\alpha a^\alpha_i$.
We may term these as the pseudo eigenvectors and  pseudo eigenvalues,  since the Hermitean matrix $i \; R$ has real
 eigenvalues $\pm \lambda^\alpha$ and real eigenfunctions $\frac{1}{\sqrt{2}} (a^\alpha_j \pm b^\alpha_j)$, and   $n \leq N/2$  is the number of  non zero eigenvalues 
  of $i \; R$. The normal form is expressed as 
$$R_{i,j}= \sum_{\alpha=1,n} \lambda^\alpha a^\alpha_i b^\alpha_j \label{rdecomp} .$$
 With this decomposition,  and with $v^\alpha = \sqrt{\lambda^\alpha} \sum_j a^\alpha_j e_j$ and
$w^\alpha = \sqrt{\lambda^\alpha} \sum_j b^\alpha_j e_j$,   we can rewrite the relation
$${\cal R}= \frac{1}{2} \sum_{\alpha=1,n} v^\alpha \wedge w^\alpha.$$
We thus see that Eq(\ref{wedge}) is possible if and only if the number of vectors $n=1$, i.e. there is only {\em one pseudo eigenvector} of $R$.
This is known as the condition of decomposability, and provides us with a neat representation Eq(\ref{rdecomp}) with a single eigenvalue $\lambda$
and a pair of orthonormal vectors $x_j \;\; y_j$
\beq
R_{i,j}= \lambda  (x_i y_j -y_i x_j). \label{rdecompose1}
\eeq
Using the  local gauge invariance in Eq(\ref{localgauge}), we can drop the condition of orthonormality of $x_j$ and $y_j$
in Eq(\ref{rdecompose1}), as far as generating solutions to the original problem  Eq(\ref{rrrr}) is concerned.
 We may also absorb the $\lambda^\alpha$ factor into the vectors, and it appears that we have $2 N$ independent real parameters in the solution 
of Eq(\ref{rrrr}). However, we observe that there is a  redundancy in this counting,  the vectors $x_j$ and $y_j$ can be changed without 
changing $R_{i,j}$ if use three linear transformations with arbitrary parameters $p,q,r$ as
\barray
(x_j,y_j) & \rightarrow & (x_j + p\  y_j,y_j) \nonumber \\
(x_j,y_j) &  \rightarrow  & (x_j, y_j + q \ x_j )\nonumber \\
(x_j, y_j) & \rightarrow & ( r \ x_j, \frac{1}{r} \ y_j). \label{scaling1}
\earray
We  thus see that the total number of real  parameters available is exactly
$2 N-3$ as known already from (I). One convenient set
of $2 N- 3$ variables  was given as $R_{1,j}$ with $2\leq j \leq N$,  and $R_{2,j}$ with $3\leq j \leq N$,  in terms of the
$x_j,y_j$ we may e.g. set $x_1=1,y_1=1,x_2=1$ and determine the remaining $2N-3$ variables from the $R_{i,j}$'s. 
 The parameterization Eq(\ref{emil})  of OWY  can be obtained  from Eq(\ref{rdecompose1}) by setting $\va_j= \frac{x_j}{y_j}$ and 
 $\gamma_j= \frac{1}{y_j}$, and the symmetries of Eq(\ref{scaling1})  are transformed into
 \barray
 (\va_j,\gamma_j) & \rightarrow & (\va_j + p,\gamma_j)  \nonumber \\
 (\va_j,\gamma_j)& \rightarrow & (\frac{\va_j}{1+ q \va_j},\frac{\gamma_j}{1 + q \va_j} ) \nonumber \\
(\va_j,\gamma_j)& \rightarrow & ( r^2 \va_j, r \gamma_j ). \label{scaling2}
\earray
We may again reduce the apparent $2N$ parameters by 3 using these relations, it amounts to choosing three parameters, say $(\va_1,\gamma_1,\gamma_2)$ arbitrarily as e.g.  $(1,1,1)$ and the rest are fixed using the inverse of Eq(\ref{emil}).

\section{Fermionic representation of commuting operators}
We next show that the matrices $Z(r)$ in \eq{z-ops} lead to a neat Fermionic representation, which may be thought of as a model for a Fermi  glass with localized states. 
 Let us define a Fermionic set of operators $a_{j},a^{\dagger}_{j}$ and $n_{j}=a^{\dagger}_{j} a_{j} $, obeying the standard anticommutation relations
\beq
\{a_{i},a^{\dagger}_{j}\} = \delta_{ij}, \label{canonical}
\eeq
with $1\leq i \leq N$.  It is elementary to see that two  commuting  matrices $ [P,Q]=0 $ lead to a commuting set of Fermionic operators (e.g. see \cite{shastry-sutherland}), i.e.
\beq
[\sum_{ij} P_{ij} a^{\dagger}_{i}a_{j}, \sum_{ij} Q_{ij} a^{\dagger}_{i}a_{j}]= \sum_{lm} [P,Q]_{lm}a^{\dagger}_{l}a_{m}=0, \label{elementary}
\eeq
where $[P,Q]$ is the matrix commutator of the two matrices $P_{ij}$ and $Q_{ij}$.  Thus we obtain a set of $N$  Fermionic operators
\beq
\hat{Z}(r) =  n_{r} + x \sum_{s}\left[ \;  \rho_{s}(r) n_{s} + S_{r,s} \left( a^{\dagger}_{r}a_{s}+a^{\dagger}_{s}a_{r} \right) \; \right]. \label{z-ops-2}
\eeq
We see that  these inherit the commutation property $[\hat{Z}(r),\hat{Z}(s)]=0 $   from \eq{z-commute}. Using the parametrization \eq{emil} and \eq{rho-par}, we write the basis set of commuting operators as
\beq
\hat{Z}(r) =  n_{r} + x \sum'_{s}\frac{1}{\va_{r}-\va_{s}} \ \left[  \gamma_{r} \gamma_{s} \left( a^{\dagger}_{r}a_{s}+a^{\dagger}_{s}a_{r} \right)  -\gamma^{2}_{r} \ n_{s} - \gamma^{2}_{s} \ n_{r}  \; \right], \label{z-ops-3}
\eeq
where the prime indicates $s\neq r$. 

 Readers   wishing  to skip the  earlier discussions, can directly verify  that the commutator $[\hat{Z}(r),\hat{Z}(s)] $ vanishes, for arbitrary values of the given parameters  by a straightforward  calculation. 

We also remark that the choice of the statistics of the canonical operators $a_{j}$ is not the only one possible. The entire argument of this section can be repeated if we use canonical Bosonic operators instead, i.e. $a_{j}\to b_{j}$ where $[b_{i},b^{\dagger}_{j}]=\delta_{ij}$. Thus one can equally well consider a Bosonic glass rather than a Fermi glass model here. 

Finally we note that the single particle sectors of the Bosonic, Fermionic and hard core Bosonic models are all identical and correspond to Type-I matrices.  For higher numbers of particles, these  correspond to other classes of matrices  depend on the statistics chosen, e.g. these  are Kronecker products of Type-I matrices in the case of canonical Fermions and Bosons.

\subsection{Mapping to The Gaudin Model }

The mapping discussed by OWY views \eq{z-ops} as the $S_{Total}^{z}=N/2- 1$ subspace representation of the Gaudin Hamiltonian\cite{gaudin,sklyanin} 
\beq
Z^{Gaudin}_{i}= S_{i}^{z} + x \sum'_{j} \frac{1}{\va_{i}-\va_{j}}\vec{S}_{i}\cdot \vec{S}_{j}. \label{gaudin}
\eeq
This model was first written down by Gaudin\cite{gaudin}. Gaudin actually  wrote it without the first term $S_{i}^{z}$,  this was supplied later by Sklyanin\cite{sklyanin} from twisting the boundary conditions. To be exact  \eq{z-ops} has an extra factor of  
$\gamma_{i}\gamma_{j}$ that OWY argue can be incorporated into the equations, and also their magnetic field term $B$ is $\propto 1/x$.  The Gaudin model  is currently very popular for describing the dynamics of Cooper pairs within the BCS theory for finite systems\cite{bcs-original, bcs}.   Each spin flip represents a Cooper pair,  from the Anderson mapping of the BCS theory to spin waves.  Thus $S^{-}_{i}=c_{k_{i}\downarrow}c_{-k_{i} \uparrow}$ and the label $i$ is actually a momentum space label. 

The point about the Fermionic representation \eq{z-ops-3} of \eq{z-ops} is that {\em it is true for all numbers of Fermions}, and not restricted to a single particle sector. In this sense, the present Fermionic representation is much more powerful, and further the factors $\gamma_{i}$ do not need any special treatment, they are automatically treated in the commutation relations. Thus  \eq{z-ops} are embedded without any further qualifications in the operator equations \eq{z-ops-3}. We see below that this representation enables us to find applications of this model for Fermions in a disordered potential, i.e. the Fermi glass problem. The same statement is also true if we use canonical Bosonic operators instead of Fermions, as mentioned above. However the Gaudin model is expressed in terms of hard core Bosons, and the magnitude of the spin is related to the $\gamma_{j}$ making the scheme somewhat cumbersome \footnote{ It is possible to include the $\gamma_{i}$
factors into an inhomogeneous 6-vertex model, provided we allow for horizontal and vertical electric fields. The only value of anisotropy that readily supports the inclusion of these fields is the Free fermi point of the 6-vertex model, so that we end up with the Fermi representation reported here. It does not seem useful to dwell on the detailed construction in view of the simplicity of the alternate argument in Eq.~(\ref{elementary}).
}.

\section{Diagonalizing the Fermi Hamiltonian and the density of states.}
We now turn to a study of \eq{z-ops-3} and a related Hamiltonian obtained by summing 
\beq
H_{R}= \sum_{r}\va_{r}\hat{Z}(r)= \sum_{r} n_{r} \va_{r} + x \sum_{i j} \gamma_{i} \gamma_{j} a^{\dagger}_{i}a_{j} - x \ \hat{N} \ \sum_{j}\gamma^{2}_{j}, \ \label{h-richardson}
\eeq
where $\hat{N}= \sum_{r} n_{r} $ is the number operator. For Fermions or canonical Bosons, this Hamiltonian is the analog of the  so called Richardson\cite{richardson} Hamiltonian in the theory of nuclear matter (the $\gamma$ factors do not usually arise in the latter). The Richardson model and also the related  BCS\cite{bcs-original} problem for finite systems\cite{bcs}, are expressed in terms of hard core bosons (i.e. spin half objects) representing Cooper pairs  $S^{-}_{i}=c_{k_{i}\downarrow}c_{-k_{i} \uparrow}$ . These are in turn, obtained by taking sums over the Gaudin $Z^{Gaudin}_{j}$ operators of \eq{gaudin}.  The  Hamiltonian \eq{h-richardson} is considerably simpler to solve for a  general  population of particles than the corresponding problem for hard core bosons, and is akin to a free gas of particles in a suitable one body potential. Clearly our Hamiltonian \eq{h-richardson} commutes with each of the $\hat{Z}(r)$, and plays a central role in the Fermi glass interpretation.

If we view the labels $i,j$ as wave vector indices, then $H$ describes a band model with an arbitrary dispersion $\va_{i} $. It is subject to  a potential that scatters from every  wave vector to each of the others, with a potential matrix element $x \gamma_{i} \gamma_{j}$.  Since  the  $\gamma_{i}$ are   arbitrary, they may be chosen at random. We thus realize a band model with a separable random scattering potential.
 If on the other hand, we view  $i$ as site labels in a tight binding model, the energies $\va_{i}$ may be chosen at  random, and the kinetic energy hops between every pair of sites- i.e.  realizing an infinite ranged random  Fermionic  Anderson model. 

We now turn to the task of diagonalizing the Hamiltonian and all the $\hat{Z}(r)$ by an orthogonal transformation.
This transformation  for the single particle sector is essentially identical to the one in Richardson \cite{richardson} ,  and many subsequent works, and hence we will be brief.
Define new canonical  Fermion set 
\beq
d^{\dagger}_{i}= \sum_{j}\ Q_{ij} \  a^{\dagger}_{j},\;\;\; \mbox{with} \;\;\; \{d_{i},d_{j}^{\dagger} \}=\delta_{ij},
\eeq
 through an orthogonal transformation generated by a real orthogonal matrix $Q$ such that  $Q^{T}\cdot Q= {\bf 1}$, and
\barray
Q_{ij}&=& \frac{\phi_{i } \  \gamma_{j}}{\omega_{i}(x) -\va_{j}}, \label{richardson-1} \\
\phi_{i}^{-2}&=& \sum_{j}\left(\frac{\gamma_{j}}{\omega_{i}(x)-\va_{j}}\right)^{2}, \label{richardson-2} \\
\frac{1}{x} &=& \sum_{j}\frac{\gamma_{j}^{2}}{\omega_{m}(x)-\va_{j}}. \label{richardson-3}
\earray
Here $\omega_{m}(x)$ in \eq{richardson-3} are the $N$ eigenvalues of the transformed Hamiltonian \eq{h-richardson}, we write the argument $(x)$  to emphasize that these 
depend parametrically on $x$.  A short further calculation gives 
\barray
H_{R}&=& \sum_{m}\omega_{m}(x) \  d^{\dagger}_{m}d_{m} - x \ \hat{N} \ \sum_{j}\gamma^{2}_{j}, \ \label{h-richardson-2}\\
\hat{Z}(j) & =& x \  \gamma_{j}^{2} \  \sum_{m}\frac{1}{\omega_{m}(x)-\va_{j}} \  d^{\dagger}_{m}d_{m}. \ \label{z-ops-4}
\earray
 As $x\to 0^{\pm}$, one sees that $\omega_{j}\to \va_{j}\pm 0$, i.e the $\omega$'s are pinned to the $\va$'s.
  The eigenvalues $\omega_{m}(x)$  are in 1-1 correspondence and  evolve out of the numbers $\va_{m}$ smoothly as $x$ increases from zero. 
Thus the  eigenvalues $\omega_{j}$ of the Richardson  Hamiltonian \eq{h-richardson} interlace the numbers $\va_{j}$, with one extremal eigenvalue that grows linearly with  $x$. For $x\gg0$ ($x\ll0$), the extremal eigenvalue $\omega_{N}\gg \va_{N}$ ($\omega_{1}\ll \va_{1}$).  The density of states of $w_{j}$ has a width that remains fixed with $x$ if we ignore the exceptional extremal case.  It is easy to see that the $\omega_{m}$ do not cross each other as $x$ varies, and they {\em do satisfy} the  von Neumann Wigner  non crossing rule.
 The conserved quantities $\hat{Z}_{j}$ may be visualized as  evolving continuously from the occupation numbers $n_{j}$ as $x$ increases from zero.

We could more generally consider the two operators formed from the sums
\beq
\hat{\alpha}(x)= \sum_{r} u_{r} \ \hat{Z}(r), \;\;\;\mbox{and}\;\;\; \hat{\beta}(x)= \sum_{r} v_{r} \ \hat{Z}(r), \label{fermi-eq1}
\eeq
with arbitrary $u_{r} $ and $v_{r}$,  and see immediately that these are the Fermi space representations of the operators introduced in \eq{eq1}and  \eq{eq2}:
\barray
\hat{\alpha}(x)&=& \sum_{r}\ u_{r} \ n_{r}+ \frac{x}{2} \sum'_{r,s}\frac{u_{r}-u_{s}}{\va_{r}-\va_{s}} \ \left[  \gamma_{r} \gamma_{s} \left( a^{\dagger}_{r}a_{s}+a^{\dagger}_{s}a_{r} \right)  -\gamma^{2}_{r} \ n_{s} - \gamma^{2}_{s} \ n_{r}  \; \right] \nn \\
\hat{\beta}(x)&=& \sum_{r}\ v_{r} \ n_{r}+ \frac{x}{2} \sum'_{r,s}\frac{v_{r}-v_{s}}{\va_{r}-\va_{s}} \ \left[  \gamma_{r} \gamma_{s} \left( a^{\dagger}_{r}a_{s}+a^{\dagger}_{s}a_{r} \right)  -\gamma^{2}_{r} \ n_{s} - \gamma^{2}_{s} \ n_{r}  \; \right] \label{eq-alpha-fermi}.
\earray
These commute mutually for any choice of the parameters, including $x$, and also with the constants of motion $Z_{j}$ in \eq{z-ops-3}, and on diagonalization become
\barray
\hat{\alpha}(x)&=& \sum_{m} \alpha_{m}(x) \ d^{\dagger}_{m}d_{m}, \nn \\
\alpha_{m}(x)&=&  x \   \  \sum_{ j }\frac{ \gamma_{j}^{2} \ u_{j}}{\omega_{m}(x)-\va_{j}} \  . \label{alpha-diag} 
\earray

A comment on the conservation laws \eq{z-ops-3} and their relationship with the ``Hamiltonian''  \eq{eq-alpha-fermi} is useful here. At $x=0$ the existence of $N$ constants of motion of the Hamiltonian is obvious since the $Z's$ are just the number operators of the Fermions.  When we perturb the Hamiltonian from this ``free case'' by adding {\em any} term proportional to $x$, we can always fix the conservation law to be valid to $O(x)$, but generally the terms do not commute to $O(x^{2})$.  This is familiar in the theory of integrability violating perturbations to integrable systems, as in the Kolmogorov- Arnold-Moser theory\cite{kam}; the conservation laws analogous to  \eq{z-ops-3} {\em can be rescued} to linear order in the new perturbations, but not to higher orders. The speciality of the specific perturbation in \eq{eq-alpha-fermi} is that there are no corrections to $O(x^{2})$ and the the conservation law \eq{z-ops-3} commute exactly. 

The inevitability of level crossings for Type-I matrices   was noted empirically in (I),  on the basis of several examples that were studied.  However  the theoretical explanation  awaited the work of OWY, who
 showed that for a generic choice of $u_{r}$, the eigenvalues of $\hat{\alpha}(x)$, i.e. $\alpha_{m}(x)$ in \eq{alpha-diag},  
 have  atleast one and at most $^{N-1}C_{2}$ level crossings as $x$ varies over its range.  These eigenvalues thus defy the  von Neumann Wigner non crossing rule,  unlike the eigenvalues of the Richardson Hamiltonian $\omega_{m}(x)$, which do  obey the  rule.  The one exceptional case is $u_{j}= \varepsilon_{j}$ when the $\alpha(x)$ reduces to the Richardson hamiltonian Eq.~(\ref{h-richardson})\footnote{In OWY, the exception is acomodated with the help of a slightly different  viewpoint. In their view Eq.~(\ref{h-richardson}) does have the requisite number of level crossings, provided   we include the limiting cases $x \to \pm \infty$. The spectrum of the large $|x|$ limit of Eq.~(\ref{h-richardson}), namely  $ \sum_{i j} \gamma_{i} \gamma_{j} a^{\dagger}_{i}a_{j} -  \ \hat{N} \ \sum_{j}\gamma^{2}_{j}, $ has a single isolated eigenvalue, and $N-1$ degenerate  (null) eigenvalues. The null eigenvalues can be viewed as consisting of $^{N-1}C_{2}$ pairwise crossings.  As $|x|$ reduces from $\infty$, some of the level crossings move towards smaller $|x|$, whereby all choices of $u_{j}$ fall into a common description. Our view is a slightly different;   at a qualitative level it seems  useful to think of the level crossing as arising from a smearing of the avoided crossing.  }.
  One may understand the violations of the non crossing rule by thinking of the eigenvalues of $\hat{\alpha}(x)$
as smeared versions  of $\omega_{m}(x)$, and thereby less sharply governed by the rule. This is illustrated in Fig.~1., where we plot the energy levels for $N=5$, and show that while $\omega_{m}(x)$ of Eq. ~(\ref{richardson-3}) avoid level crossings, the derived eigenvalues $\alpha_{m}(x)$ from Eq.~(\ref{fermi-eq1} ,\ref{alpha-diag})  {\em do display level crossings}.   In this sense, there is  a hidden generic model $H_{R }$ satisfying the non crossing rule, behind the violations of the same in the constructed matrices $\alpha({x})$.
 \begin{figure}[!tbh]
\includegraphics[width=7.5cm]{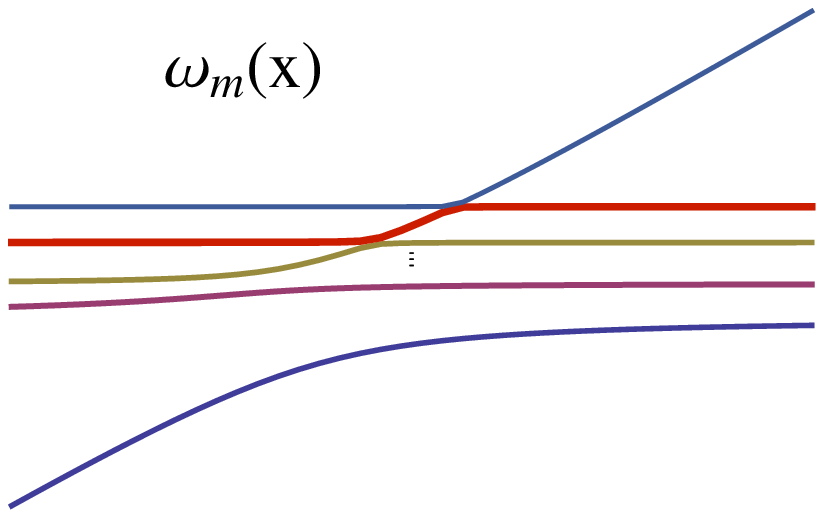}
\includegraphics[width=7.5cm]{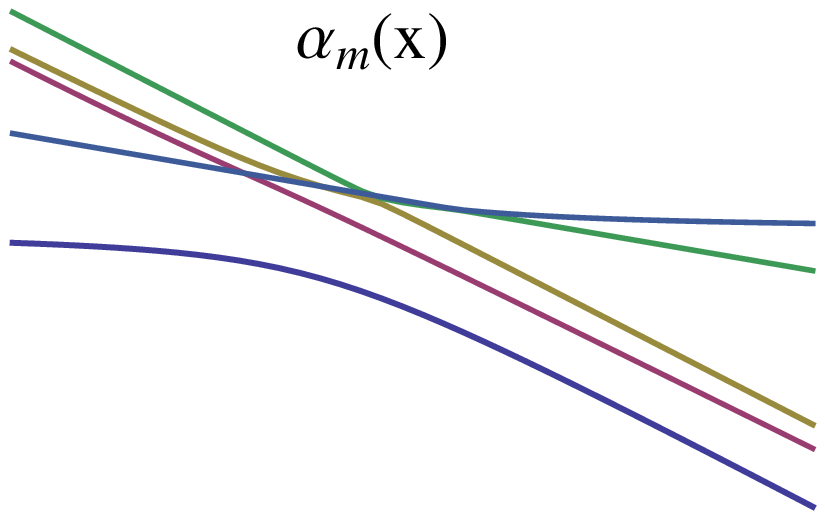}
\caption{Left panel shows the eigenvalues $\omega_m(x)$ from Eq.~(\ref{richardson-3}) for the case of $N=5$, where one observes narrowly avoided level crossings involving  the  top three levels. Right panel shows the effect of mixing levels through Eq.~(\ref{fermi-eq1} ,\ref{alpha-diag}) with  $u_{j}- \varepsilon_{j} $ chosen randomly with a small scale of variation. We see that the  eigenvalues $\alpha_{m}(x)$  obtained from Eq.~(\ref{fermi-eq1},\ref{alpha-diag}),  cross each other profusely, thereby violating the Wigner von Neumann non crossing rule.
}
\label{figure}
\end{figure}

Finally, we note that the Hamiltonians \eq{eq-alpha-fermi} with Fermionic $\alpha_{j}$  (Bosonic $b_{j}$) can be viewed as representing a class of localized states in the Fermi (Bose) glass problem of disordered non interacting quantum particles. At $x=0$, the model consists of localized states with energies $u_{r}$, and clearly has $N$ conservation laws $Z(r)$ as in Eq.~(\ref{z-ops-3}), corresponding to the occupation numbers of the different sites.  As $x$ varies from zero, the particles hop around as dictated by the Hamiltonian, but  with generalized conserved occupancies at all  sites given by \eq{z-ops-3}. These are therefore  localized to all orders in the perturbation $x$, despite hoppings that carry them far away.  We can easily see that the energy level statistics of these systems follow the Poisson distribution for small separations, due to the   level crossings that occur in these Hamiltonians. The absence of level repulsion  what   one expects from localized states in the Anderson model on general grounds.

\section{Acknowledgements}
 I thank    M. S.  Narasimhan and    T. R. Ramadas  for   valuable  discussions regarding the Pl\"ucker relations,   Emil Yuzbashyan for a stimulating correspondence, and    H. Haber for helpful comments.  I thank  the Raman Research Institute (Bangalore, India) for hospitality, where a part of the manuscript was written. This work was   supported by DOE through  a grant BES  DE-FG02-06ER46319.


\end{document}